\documentclass[twocolumn, amsmath]{revtex4-1} 
\usepackage{graphicx}% Include figure filesf
\usepackage{epstopdf}
\usepackage{epsfig}
\usepackage{bm}% bold math
\usepackage{subfigure} % for combining figures
\begin{document}

\title{On the critical behavior of the 
Susceptible-Infected-Recovered (SIR)
model on a square lattice}
\author{T\^ania Tom\'e$^1$ and Robert M. Ziff$^2$ \\
$^1$Instituto de F\'{\i}sica, Universidade de S\~ao Paulo \\
Caixa Postal 66138, 05315-970 S\~ao Paulo SP, Brazil \\
and \\
$^2$Michigan Center for Theoretical Physics \\
and Department of Chemical Engineering, University of Michigan \\
Ann Arbor, MI 48109-2136, USA }

\begin{abstract}
By means of numerical simulations and epidemic analysis, 
the transition point 
of the stochastic, asynchronous
Susceptible-Infected-Recovered (SIR) model on a square lattice
is found to be $c_{0}=0.1765005(10)$, where $c$ is the probability a chosen infected
site spontaneously recovers rather than tries to infect one neighbor.
This point corresponds to an infection/recovery rate of $\lambda_c = (1-c_0)/c_0 =  4.66571(3)$
and a net transmissibility of $(1-c_0)/(1 + 3 c_0) = 0.538410(2)$, which falls
between the rigorous bounds of the site and bond thresholds.
The critical behavior of the
model is consistent with the 2-d percolation universality class,
but local growth probabilities differ
from those of dynamic percolation cluster growth, as is demonstrated 
explicitly.

\end{abstract}

\maketitle

%--------------------------------------------------
\section{Introduction}

The Susceptible-Infected-Recovered (SIR) model is a fundamental system in
epidemiological modeling 
\cite{Kermack27,Bailey53,Dietz67,Mollison77,Renshaw91,Hastings96,
Murray03,Keeling08}
and has been studied extensively within the context of
non-equilibrium phase transitions and critical phenomena
(i.e., \cite{Grassberger83,CardyGrassberger85,Hinrichsen00,Antal01a,Newman02,
SanderWarrenSokolov03,DammerHinrichsen03,DammerHinrichsen04,JooLebowitz04,ArashiroTome07,deSouzaTome10,
Henkel08,SchutzBrandauTrimper08,BlackMcKaneNunesParisi09,AssisCopelli09}).
The model was developed to describe the propagation of an epidemic that
occurs during a period of time much smaller than the lifetime of individuals
of a given population. It is assumed that the population can be divided into
three categories: Susceptible (S), Infected (I) and Recovered (R) individuals.
Susceptible individuals become infected at a given rate through contact with
infected individuals. Infected individuals recover with a given rate and
become immune and recovered. The model is capable of showing a threshold of
the epidemic spreading as one increases the infection rate.

The SIR process has been studied using different approaches and contexts.
Originally it was defined in 1927 by Kermack and McKendrick \cite{Kermack27}
as a deterministic process by means of a set
of ordinary differential equations; they showed that epidemics
disappear before all the susceptible individuals contract the disease.
Afterwards the model was given a stochastic description
by means of birth and death processes  \cite{Bailey53},
and later
Grassberger \cite{Grassberger83} introduced a cellular automaton implementation,
that is, a synchronous-update Markovian process on a lattice. 
In this paper we consider a stochastic, asynchronous-update lattice version of 
the SIR model  \cite{deSouzaTome10}, 
in which lattice sites are updated one at a time.
This model is a special case of the predator-prey stochastic lattice-gas
model introduced by Satulovky and Tom\'e \cite{Satulovsky94,Satulovsky97}, 
and also considered by Antal et al.\ \cite{Antal01a,Antal01}.
It is also a special case of the Susceptible-Infected-Removed-Susceptible (SIRS) 
stochastic lattice gas model \cite{deSouzaTome10}.
For the synchronous or asynchronous versions 
of the SIR process, one observes
that, as the model parameters are varied, a phase transition takes place.
This is a continuous phase transition between two distinct regimes: one in
which the population remains susceptible (non-spreading regime) and another in
which the epidemic spreads over the lattice (spreading regime), 
where a significant portion of 
the population becomes infected and eventually immune. At the
transition the system becomes critical and corresponds to the epidemic
threshold. Cardy and Grassberger \cite{CardyGrassberger85} argued that the
transition found in the SIR cellular automaton is of the percolation
universality class. This has been confirmed in various ways by numerous studies
in stochastic models with synchronous as well as asynchronous update 
\cite{Hinrichsen00,Antal01a,DammerHinrichsen03,DammerHinrichsen04,ArashiroTome07,deSouzaTome10}.

In recent years there has been a great deal of interest
in the SIR model on networks and other systems
(i.e., \cite{Newman02,SanderWarrenSokolov03,DammerHinrichsen03,DammerHinrichsen04,JooLebowitz04,ArashiroTome07,deSouzaTome10,
BenNaimKrapivsky04,Hastings06,SerranoBoguna06,KesslerShnerb07,AlonsoMcKanePascual07,KenahRobins07,Miller07,Volz08,SchutzBrandauTrimper08,AssisCopelli09,BlackMcKaneNunesParisi09,TomedeOliveira10}).  
In this paper we consider the SIR model on the square lattice, which has 
wide applications to many physical problems such as the
spread of disease in plants, and which has been studied
only to a limited extent \cite{Kuulasmaa82,Grassberger83,SanderWarrenSokolov03,deSouzaTome10}.
One question that has come up in the context of mean field and network studies
is the effect of heterogeneity in
the infectious period \cite{KenahRobins07,Miller07,Volz08}.
When the period of infection is fixed,
there is a direct mapping of the model to bond percolation
on the lattice, and thus the transition point can be determined
exactly \cite{Kuulasmaa82,Grassberger83,KuulasmaaZachary84,SanderWarrenSokolov03}.
However, when the infection period
is heterogeneous, such as the exponentially distributed
infection period inherent to the original SIR model, there is no
exact solution.
In this paper, we carry out a careful numerical
study of the transition point for the exponentially distributed  case, and then investigate the
correlations in the transmission of the
infection to the nearest neighbors.  We work out the correlations explicitly,
and find that at the critical point there is a higher probability to infect more
rather than fewer of the neighbors, unlike the case of bond percolation
where the distribution of infected neighbors is simply binomial with $p = 1/2$.

The rest of the paper is organized as follows:
In section \ref{sec:master} we define the asynchronous SIR 
model in terms of the master equation. In section \ref{sec:simulations}
we describe the Monte-Carlo algorithm that we use, and 
in section \ref{sec:numerical} we carry out a numerical analysis
of the cluster size distribution of recovered individuals
 to precisely determine the critical
point of the model.
In section \ref{sec:perc} we show in detail how the asynchronous SIR model
differs from percolation on a local scale.  In section \ref{sec:other}
we discuss the relation of our results to some other population biology models,
and in section \ref{sec:conclusions}
we give our conclusions.

%-------------------------------------------------
\section{The SIR lattice model}
\label{sec:master}

To model the dynamics of an epidemic with immunization we consider a
stochastic lattice-gas model with asynchronous dynamics. 
The lattice plays the role
of the spatial region occupied by the individuals and the lattice sites are
the possible locations for the individuals. 
Each site can be occupied by just one individual
that can be either a susceptible,
an infected, or a recovered individual,
called, respectively, an S site, I site and R site.
At each time step a site is
randomly chosen and the following rules are applied: 
(i) If the chosen site is in state S or R it remains unchanged.
(ii) If it is in state I then (a) with probability $c$ the
chosen site becomes R and (b) with the complementary probability
$b=1-c$ a neighboring site is chosen at random. If the chosen
neighboring site is in state S it becomes I; otherwise it remains
unchanged. Notice that a site in state R remains forever
in this state so that the allowed transitions are S$\to$I$\to$R.
From this set of dynamic rules 
it follows that the state of the system will change
only when the chosen site is in state I, a feature that will be used to speed
up the simulation as explained in section \ref{sec:simulations}.
This algorithm is equivalent to making S$\to$I with probability
$b n_\mathrm I / 4$, where $n_\mathrm I$ is the number of nearest-neighbor I sites,
and I$\to$R with probability $c$.

The system evolves in time according to a master equation. 
To each site $i$ of a two-dimensional lattice,
we associate a stochastic variable $\eta_{i}$
that takes the values $0$, $1$, or $2$ 
according to whether the site $i$ is occupied by 
an R or an S or an I individual, respectively.
A microscopic configuration of the entire system is denoted by the
stochastic vector $\eta =(\eta_1,\dots,\eta_i,\ldots,\eta_N)$ where $N$ is
the total number of sites. Because the possible transitions are the cyclic
ones ($1\to2\to0$), it is convenient to define the state $\eta^i$ obtained
by a cyclic permutation of the state of site $i$, that is, 
$\eta^i=(\eta_1,\dots,\eta_i^\prime,\ldots,\eta_N)$ where $\eta_i^\prime$ is 
$1$, $2$, or $0$ according to whether $\eta_i$ is $0$, $1$, or $2$,
respectively. 
The master equation for the probability distribution $P(\eta,t)$ associated
with the microscopic configuration $\eta$ at time $t$, is given by 
\[
\frac{d}{dt}P(\eta,t)=
\sum_j{\sum_i}^\prime\{w_{ij}^B(^i\eta)P(^i\eta)-w_{ij}^B(\eta )P(\eta )\}+
\]
\begin{equation}
+\sum_j\{w_j^C(^j\eta)P(^j\eta) - w_j^C(\eta)P(\eta) \}\ ,
\label{2}
\end{equation}
where the summation on $i$ extends over the nearest-neighbor sites
of site $j$ and $^j\eta$ denotes the state obtained from $\eta$
by an \emph{anticyclic} permutation of state (0$\to$2$\to$1).
The quantity $w_{ij}^B$ is 
the transition rate associated with the infection process, 
given by 
\begin{equation}
w_{ij}^B(\eta)=\frac{\beta}{{z}}\delta(\eta_{i},1)\delta(\eta_{j},2) \ ,
\end{equation}
where ${z}$ is the number of nearest-neighboring sites
of site $j$ (the lattice coordination number)
and $\delta(x,y)$ is the Kronecker delta; and  
$w_j^C$ is the transition rate for the recovery process,
given by
\begin{equation} 
w_j^C(\eta) = \gamma \delta(\eta_j,2) \ .
\end{equation}
Two external parameters are
associated to these processes: the infection rate $\beta$ and the recovery 
rate $\gamma$. The probabilities $b$ and $c$ are related to these
rates by $b=\beta/(\beta+\gamma)$ and $c=\gamma/(\beta+\gamma)$
so that 
\begin{equation}
b+c=1 \ ,
\label{21}
\end{equation}
as it should.

From the master equation (\ref{2}) 
we can derive the time evolution equations
for the densities of recovered $\rho_{0}$, 
susceptible $\rho_{1}$, and infected $\rho_{2}$.
The connection of the present stochastic lattice
model with the approach developed by 
Kermack and McKendrick \cite{Kermack27} can be revealed
by using a simple mean-field approximation.
Within this approximation \cite{deSouzaTome10} 
the following set of ordinary differential
equations for the densities can be derived
\begin{equation}
\frac{d}{dt}\rho_1 = -\beta \rho_1\rho_2 \ ,
\label{eq:meanfield1}
\end{equation}
\begin{equation}
\frac{d}{dt}\rho_2 = \beta \rho_1\rho_2 - \gamma \rho_2 \ ,
\label{eq:meanfield2}
\end{equation}
\begin{equation}
\frac{d}{dt}\rho_0 = \gamma \,\rho_2 \ .
\label{eq:meanfield3}
\end{equation}
These equations are essentially the equations introduced
by  Kermack and McKendrick \cite{Kermack27}
in their deterministic approach for the spreading of
a disease with immunization.

To analyze an epidemic, one can begin from an initial condition
at time $t=0$ where all the individuals are susceptible, with 
the exception of a very small number of infected individuals; 
that is, $\rho_1=1-\rho^*$, $\rho_2=\rho^*\ll 1$ and $\rho_0=0$.
Using this initial condition and Eqs.\ (\ref{eq:meanfield1}-\ref{eq:meanfield3}), one finds that 
the system evolves in time and reaches two types of states: 
one where the epidemic spreads, that is, 
$\rho_{0}\neq 0$, $\rho_{1}\neq 0$, $\rho_{2}=0$, when $t\to\infty$,
which  occurs for sufficiently large values of the infection 
probability $b$,
and  another where the epidemic does not spread, that is, 
$\rho_{1}=1$ when $t\to\infty$, which occurs for small values of $b$.
As one varies the parameter $b$, there is a continuous 
phase transition at $b=1/2$. 
In stochastic lattice models one expects similar behavior
but a distinct value for the critical parameters.
In that case one starts with a lattice full of susceptible
individuals with the exception of a single infected individual,
and studies the epidemics that ensue.

%--------------------------------------------------
\section{Simulation algorithm}
\label{sec:simulations}

The asynchronous SIR 
model may be simulated by a kinetic Monte-Carlo process
by following procedure:

\begin{itemize}

\item Pick an I site randomly from a list of all I sites. 

\item Generate a random number $X$ in $(0,1)$.  
If $X \le c$ then let I $\to$ R.

\item Otherwise (if $X > c$), pick one nearest neighbor 
of the I site randomly. 
If the neighbor is S, then make it I and add to list of I sites.

\item Repeat as long as there are available I sites.

\end{itemize}

When we remove an I site from the list in the first step, we
swap the empty location with the site at the top of the list and decrease
the list length by one in order
to keep the list compact.  New I sites in the third step are added
to the top of the list. 

The Monte-Carlo time $t$ is determined by incrementing $t$
by $1/N_\mathrm I$, where $N_\mathrm I$ is
the current number of I sites on the list, each time an I is picked
from the list.  This reflects an effective increase of one Monte-Carlo
step if every site on the lattice were chosen once, on the average.  
However, as explained below, we don't actually keep track of
time in our simulations, but instead monitor just the size
(number of R sites) of the clusters.

This model is closely related to standard percolation growth, in which an
active growth site spreads to an unoccupied nearest neighbor with
the given bond probability. However, in
the percolation case, as described below, the growth site becomes 
inactive after the
four nearest neighbors are checked (for both asynchronous and synchronous
updating). In the SIR model, the I site remains active an exponentially 
distributed
length of time, and can thus attempt to infect neighboring sites multiple times. That
behavior leads to a kind of correlation in the spreading in the SIR model,
as we shall see.

In order to study properties of this model in detail, it is necessary to
know the transition point to high accuracy. Recent work \cite{deSouzaTome10}
has shown that $c_{0}=0.1765(5)$, where the number in parentheses represents
the error in the last digit. In this, paper, we find $c_0$ to about 200
times the accuracy, and indeed, in so doing, we find nearly complete 
agreement with the previous central value.

%--------------------------------------------------
\section{Numerical results}
\label{sec:numerical}

To find the transition point, we consider the statistics of the 
cluster size distribution, and determine 
$c_0$ as the point where the distribution follows a power law. 
According to standard
percolation theory, the probability $P_s$ that a point belongs to a cluster
containing $s$ sites is given by $s n_s$, where $n_s$ is the number of 
clusters of
size $s$ (per site) on the lattice. At criticality, $n_s \sim s^{-\tau}$, so
that $P_s \sim s^{1 - \tau}$. Integrating from $s$ to $\infty$ we find the
probability $P_{\ge s}$ that a point belongs to a cluster of size greater or
equal to $s$. At criticality, $P_{\ge s} \sim s^{2 - \tau}$, and within the
scaling region one expects \cite{StaufferAharony94}
\begin{equation}
P_{\ge s} \sim s^{2 - \tau} F(\varepsilon s^\sigma) \approx s^{2 - \tau} 
(A + B \varepsilon s^\sigma )   \ ,
\label{eq:Pges}
\end{equation}
where $\varepsilon=c - c_0$ and
for the last term we made a Taylor-series expansion of the scaling
function $F$ for small values of its argument \cite{LorenzZiff98}. 
In 2-d percolation, we have 
$\tau = 187/91$ and $\sigma = 36/91$. While there is no ``$n_s$" for dynamic
percolation (because we don't have a lattice fully populated with clusters),
we expect $P_{\ge s}$ for dynamic percolation and the SIR model at criticality
to show the same power-law behavior as static percolation.

\begin{figure}[thbp]
\includegraphics[width=3in]{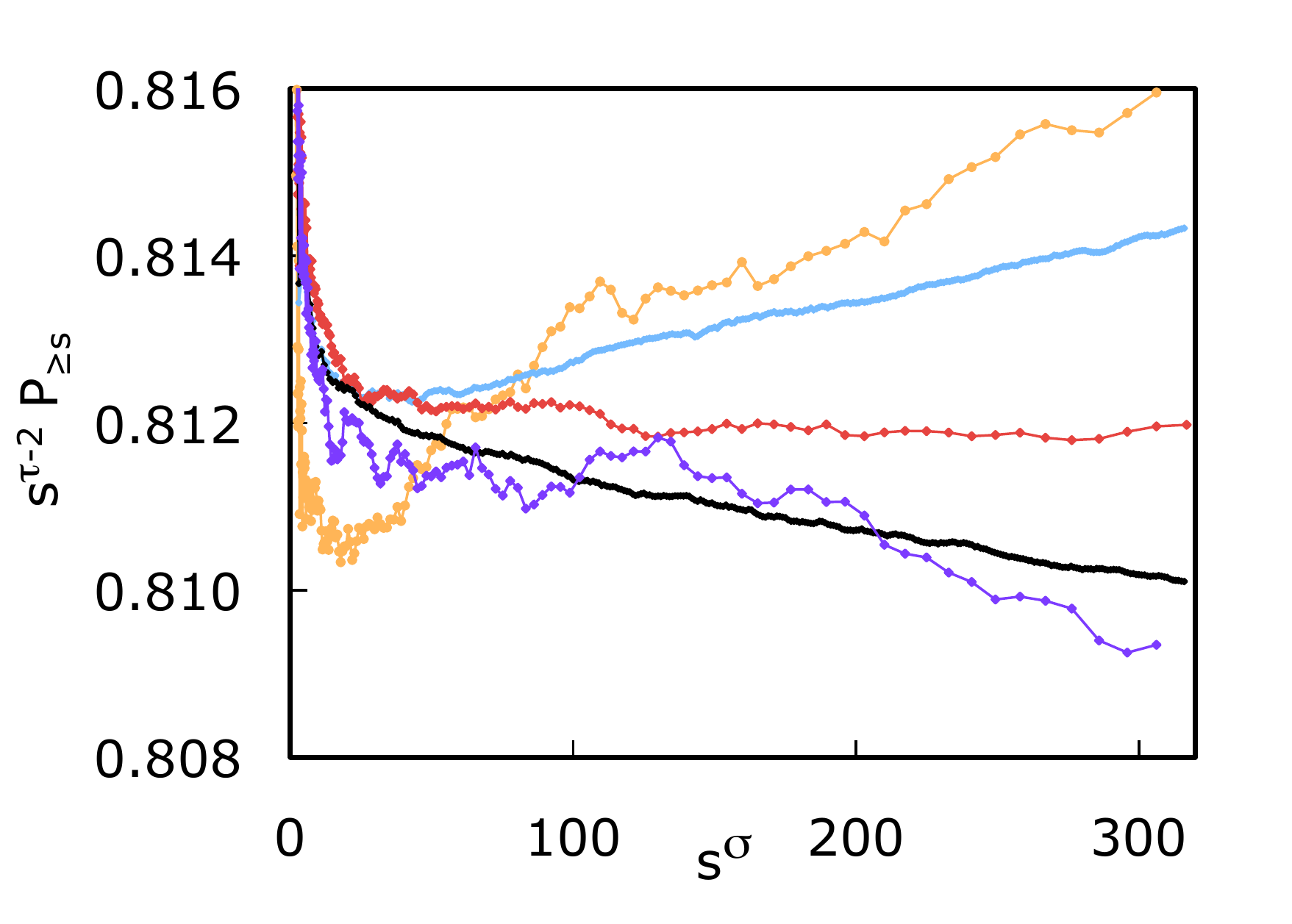} \vspace{.2in}
\caption{(Color online) Plot of $s^{\tau - 2} P_{\ge s}$ vs.\ $s^{\sigma}$ for the asynchronous SIR model,
using dynamic percolation values $\tau = 187/91$ and $\sigma = 36/91$, 
for  (top to bottom for large $s$): 
$c = 0.176490$, $0.176495$, $0,176500$, $0.176505$, and $0.176510$.  
The number of samples are, respectively, 
335000, 10500000, 440000, 11000000, 360000.}
\label{fig:SIRplot}\centering
\end{figure}

To calculate  $P_{\ge s}$, we ran simulations with the lattice covered 
entirely by S sites, except for
a single I site in the center, using the algorithm above. The lattice size
was 16384 $\times$ 16384. The size $s$ of the cluster was characterized by
the number of R sites. When $s$ went beyond a cut-off of
$2^{21}=2097152$, the growth was stopped. The cluster sizes were binned
to make a histogram. Various runs were made at each value of $c$
as listed in the caption of Fig.\ \ref{fig:SIRplot}.  
For a pair of values of $c$ which
bracketed the apparent transition point, we generated about $10^7$ clusters,
requiring several weeks of computer time each.  The random number
generator we used was R(9689) of \cite{Ziff98}.
 
\begin{figure}[htbp]
\includegraphics[width=3in]{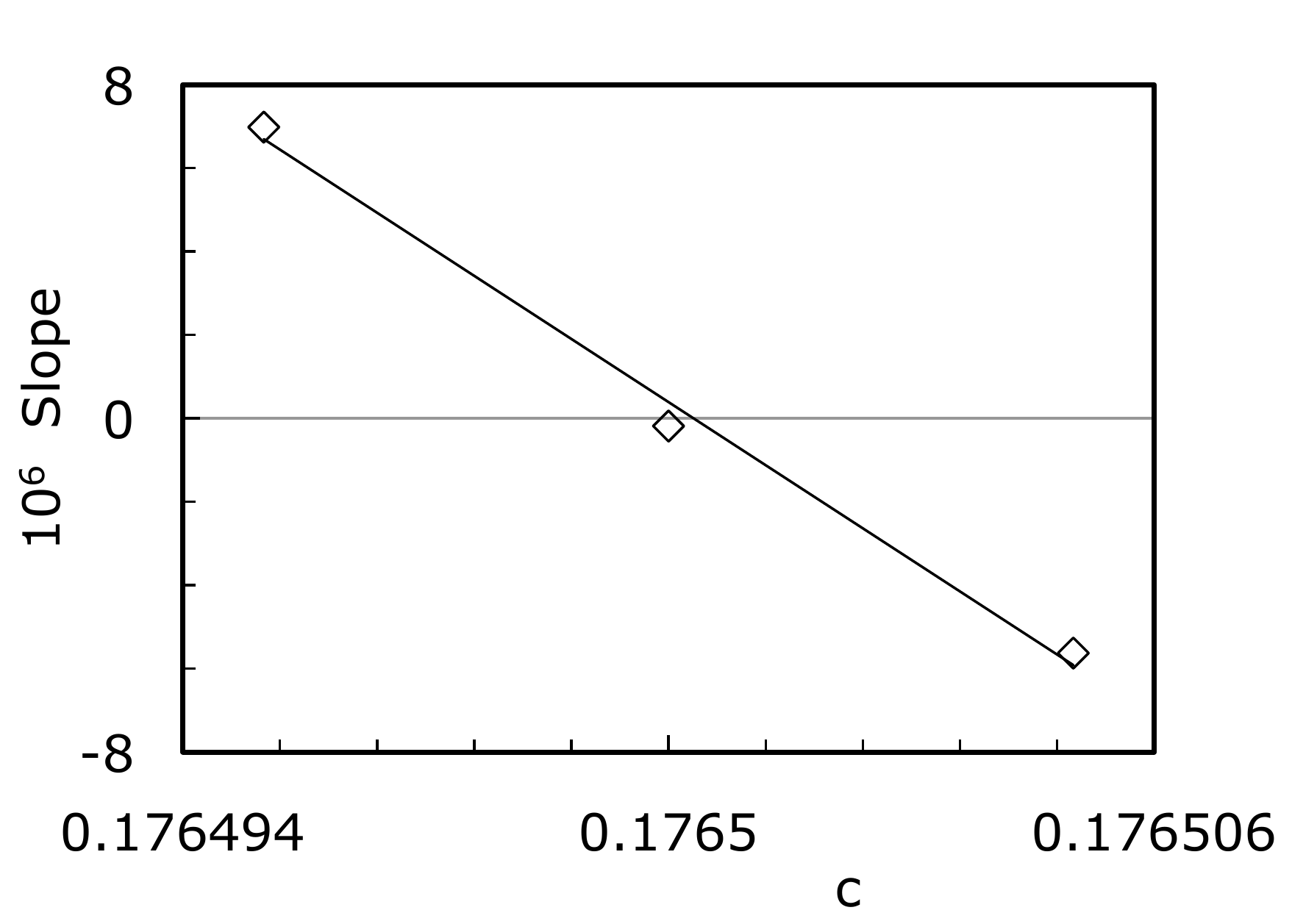} \vspace{.2in}
\caption{Slope of the linear part of the three central curves of 
Fig.\ \ref{fig:SIRplot}, measured for $s^\sigma \ge 125.$  The 
intercept of the line on the horizontal axis gives an estimate for $c_0$.  
}
\label{fig:SIRslope}
\centering
\end{figure}

\begin{figure}[thbp]
\includegraphics[width=3in]{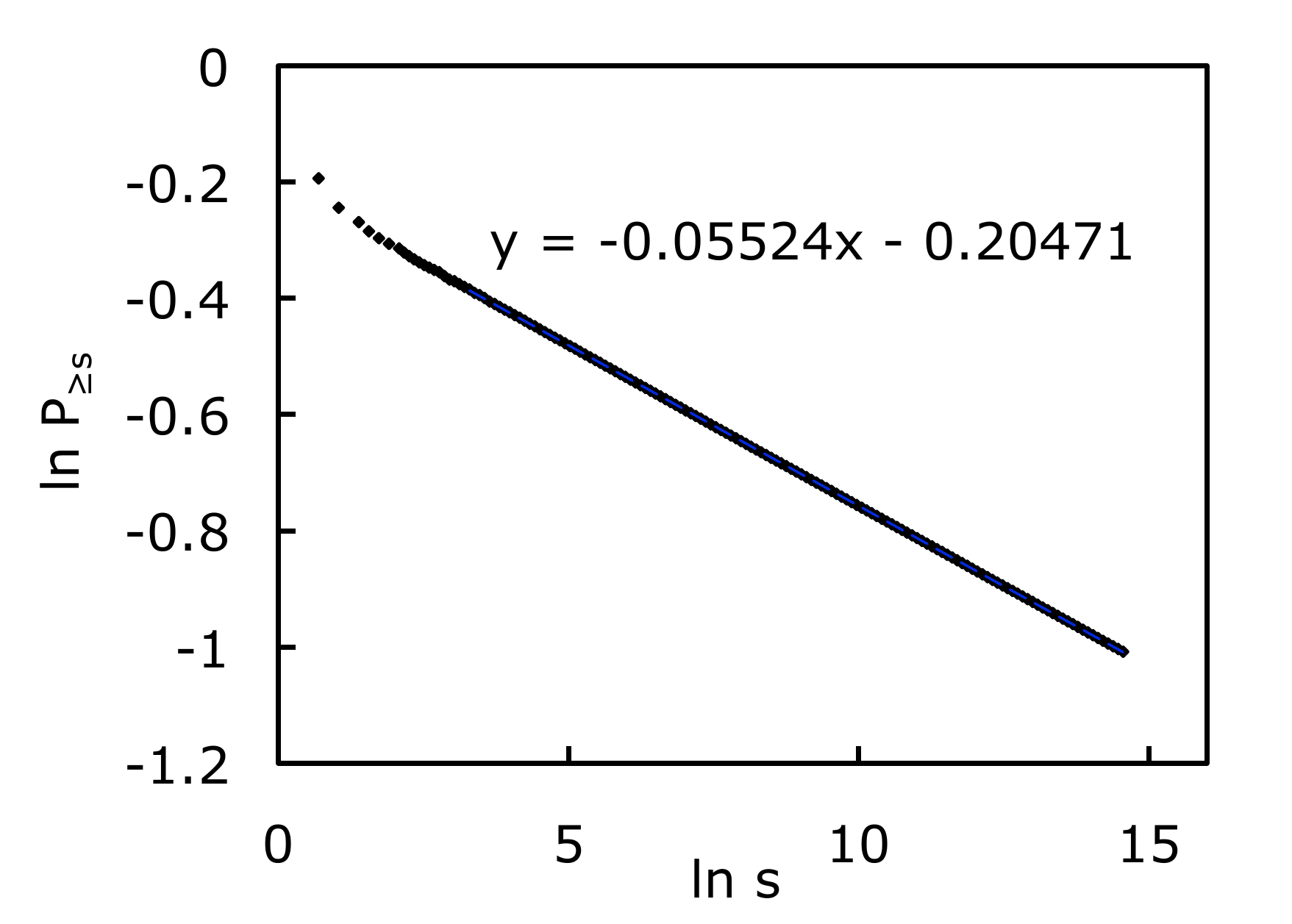} \vspace{.2in}
\caption{Plot of $\ln P_{\ge s}$ vs.\ $\ln s$ for the SIR model,
simulated on a $16384 \times 16374$ lattice, with a cutoff
of $s = 2097152$.
The equation represents a linear fit of the points for $s > 100$,
where $x$ represents $\ln s$ and $y$ represents $\ln P_{\ge s}$.}
\label{fig:percSIRslopeplot}\centering
\end{figure}

While the time $t$ might perhaps be more of a natural variable to consider 
for dynamic percolation, we chose to consider the survival probability as a function 
of $s$.
One advantage of using $s$ instead of $t$ is that the percolation exponents 
for $P_{\ge s}$ are known exactly in two dimensions, while those for $P_{\ge t} \sim t^{(2 - \tau) D/d_\mathrm{min}} = t^{-0.09213\ldots}$ (at criticality) 
are related 
to $d_\mathrm{min}$ which is known only approximately:
$d_\mathrm{min} \approx 1.1306(3)$ \cite{Grassberger99}.
(Here, $D=91/48$ is the fractal dimension.) 
Also, it is somewhat easier to keep track of the dependence upon a discrete variable 
$s$ rather than a real variable $t$.
In Fig.\ \ref{fig:SIRplot} we plot $s^{\tau - 2}P_{\ge s}$ vs. $s^\sigma$
for various $c$, using the 2-d
percolation values of $\tau$ and $\sigma$.  According to (\ref{eq:Pges}),
this plot should be a linear function, with a slope proportional to $c - c_0$,
and slope equal to zero for $c = c_0$. Indeed, this linear behavior is followed
very well, except for smaller $s$ where (\ref{eq:Pges}) is not valid due
to finite-size effects. 
in Fig.\ \ref{fig:SIRslope}, we plot the slopes of the linear parts
of the three central values as a function of $c$.  The intercept
where the slope is zero gives the critical point:
\begin{equation}
c_0 = 0.1765005(10)
\end{equation}
where the error is based upon the statistical error of the data.
The linear behavior of the curves in Fig.\ \ref{fig:SIRplot} for larger $s$ 
shows that the critical behavior is consistent with percolation scaling.  
In Fig.\ \ref{fig:percSIRslopeplot}, we plot 
$\ln P_{\ge s}$ vs.\ $\ln s$ at $c = 0.1765$, 
and find a slope $-0.05524$ consistent with the percolation 
prediction $2 - \tau = -5/91 \approx 0.054945$.

We mention that we also confirmed that $c_0 \approx 0.1765$ is
the transition point for the alternate simulation method (as used in
\cite{deSouzaTome10}) in which 
all sites are sampled, and I$\to$R with probability $c$ and 
S$\to$I with probability $n_\mathrm I (1-c)/4$.  This simulation runs more slowly because
of the time spent testing S sites with no I nearest-neighbors
(although it can be speeded up by using a list of eligible S sites).
The program can also be speeded up somewhat by raising both probabilities the maximum amount:
I$\to$R with probability $c' = c/(1-c)$ and 
S$\to$I with probability $n_\mathrm I/4$.  We also confirmed by simulation that this 
procedure yields the correct critical point $c' = c_0/(1 - c_0) \approx 0.21433$.

In terms of the probability of infection (or the net \emph{transmissibility}) our result
implies by (\ref{eq:probinfected}),  prob(infection) $ = 0.538410(2)$,
which falls within the rigorous lower and upper bounds \cite{Kuulasmaa82,Grassberger83}
of $p_c^\mathrm{(bond)} = 1/2$ and $p_c^\mathrm{(site)} \approx 0.592746$.

%-------------------------------------------------
\section{Relation to percolation growth}
\label{sec:perc}

While the asynchronous SIR model is clearly in the dynamic percolation
universality class, it is not strictly identical to percolation.  
In this section
we show explicitly how the two differ locally by considering the probabilities
of an infection spreading from a single I site.  This difference between
the SIR model and bond percolation was noted by Kuulasmaa \cite{Kuulasmaa82}
and by Grassberger \cite{Grassberger83}, and more recently discussed
in general by Kenah and Robins \cite{KenahRobins07}.  Here we carry out
a brief analysis directly related to the computer algorithm we developed, and
give explicit numerical results for the infectivity or transmissibility at the
critical point. 

First, for comparison, we formulate epidemic bond percolation growth in the SIR language.
As in the algorithm in section \ref{sec:simulations}, we start with a system of all S, with one I site
at the center.  The algorithm we follow for percolation is:

\begin{itemize}

\item Pick an I site randomly from the list of I sites.  Set this I to R and remove it from the list.

\item Consider the I site's four nearest-neighbor sites, and for each do the following:

\item  If the neighbor is an S site, then generate a random number $X$ 
in $(0,1)$

\item If $X \le p$ then let the S become I, and add to list;
otherwise, do nothing.

\item Repeat as long as there are available I sites.

\end{itemize}

This is equivalent to bond percolation because each possible bond is considered
only once and with a fixed probability $p$.  Note that the algorithm does not generate
all the bonds of a cluster, as no internal bonds are considered, but it finds
all the ``wetted sites" with the correct probability.  The bonds that are 
occupied form a minimum spanning tree on the cluster.  A similar algorithm
was given by Grassberger \cite{Grassberger83}.

If a given I site has $m$ nearest-neighbor S sites ($m = 0, \ldots, 4$), 
then the probability
$P_k^{(m)}$ that exactly $k$ of them become infected as a result of the central I site 
for percolation is just 
\begin{equation}
P_k^{(m)} =\binom{m}{k} p^k (1-p)^{m-k}  \ .
\label{eq:binomial}
\end{equation}
At the threshold $p_c = 1/2$, for $m = 4$, for example,
we have $P_0^{(4)} = 1/16$, $P_1^{(4)} = 4/16$, $P_2^{(4)} = 6/16$, 
$P_3^{(4)} = 4/16$ and $P_4^{(4)} = 1/16$, with
an average occupancy of 2.

Next we calculate $P_k^{(m)}$ for the SIR model.  First, we derive the probability of spreading
for ``transmissivity" for the SIR model.
Consider an I site surrounded by at least one S site.
We desire the probability that a given one of those S sites will become infected
by the I site.
The I site will remain infectious for an exponentially distributed number of trials.
The probability $p(n)$ that it remains infectious for 
$n$ trials is simply
\begin{equation}
p(n) = (1-c)^n c \quad n = 0, 1, 2,\ldots  \ ,
\label{eq:pn}
\end{equation}
because $c$ is
the probability that it recovers in a given trial.  
This implies that, on the average, an I site will
be considered $\sum_{n=0}^\infty n p(n) = (1-c)/c$ times before it recovers.

For each trial that occurs on the given I site,
we check one of its four neighbors randomly to see if any is S,
not remembering if we have already checked that neighbor before.
The probability that a given nearest neighbor S
site is chosen in at least one of the $n$ trials is given by $1 - (3/4)^n$.
Multiplying by $p(n)$ from Eq.\ (\ref{eq:pn}) and summing, we find the net probability a given
nearest-neighbor S site becomes infected:
\begin{eqnarray}
\mathrm{Prob(infected)}&=&\sum_{n=0}^\infty (1-c)^n c 
\left[1 - \left(\frac{3}{4}\right)^n\right] \nonumber \\
& =& \frac{1-c}{1 + 3 c}
\label{eq:probinfected}
\end{eqnarray}
If we set this probability equal to $1/2$ as in standard bond percolation, we
would find $c = 1/5$, which is above the observed value $c_0 = 0.1765$.
Thus, finding where the effective bond probability equals $1/2$ does \emph{not} give
the correct threshold.  This is because that bond is not occupied
independently, but is correlated with its neighbors, and so the effective
bond probability (probability of spreading to a neighbor) is not 1/2.  
Putting $c = 0.1765$ into this formula, we find Prob(infected) $= 0.5384$,
which is greater than in standard bond percolation.  

In another limit, the SIR model becomes identical
to site percolation on the square lattice  \cite{Grassberger83}.
Here, one assumes that if an I site does not recover in its
first trial, all neighboring S sites become infected.
The threshold corresponds to $1 - p(0) = 0.592746 = p_c^\mathrm{(site)}$,
or $c = 1 - p_c^\mathrm{(site)} = 0.407254$,
where $p_c^\mathrm{(site)}$ is the site percolation threshold on the square lattice
\cite{NewmanZiff01,Lee08,FengDengBlote08}, and have used $p(0)$ from Eq.\ (\ref{eq:pn}). This also implies
Prob(infected) $= 0.592746$.  Thus, we 
see that the net rate of infection of the SIR model (0.5384) 
falls between  that of the (uncorrelated) bond percolation
and the highly correlated site percolation values, as mentioned earlier.

Another comparison to make is to the SIR model defined on the Bethe lattice (Cayley tree).  
Say that we have a Bethe lattice with coordination number ${z}$.  Then an extension of equation
(\ref{eq:probinfected}) to arbitrary ${z}$ gives Prob(infected) $= b / [{z} - ({z}-1) b]$.
Setting this equal to $1/({z}-1)$, because an I site has $({z}-1)$ nearest-neighbor S sites and the critical point is when, on the average, one nearest-neighbor site is infected,
we find \cite{TomedeOliveira10}
\begin{equation}
b = \frac{{z}}{2 ({z}-1)}   \ .
\end{equation}
  Thus, for ${z} = 3$, the transition is at $c = 1/4$, and for ${z} = 4$, the transition is at  $c = 1/3$.  The latter is much greater than the critical $c_0$ we found for the square lattice.  This is expected, because for the Bethe lattice the net bond probability should be less than for a regular lattice with the same $z$.

Now, we turn to the probabilities $P_k^{(m)}$ for the SIR model.  
From a recursion relation analysis, and using the generating
function method, we find the coefficients for each $m = 1, 2, 3, 4$.
The derivation is given in the Appendix. The resulting $P_k^{(m)}$ calculated from Eqs.\ (\ref{eq:recursion1}-\ref{eq:recursion2}) in the Appendix
at $c = 0.1765$ are given in Table \ref{table:SIR}.  
The distribution is seen to be much
different from a binomial distribution of standard percolation Eq.\ (\ref{eq:binomial}),
with an increase in the probability
with increasing $k$ for a given $m$.  
The probabilities add to $1$, and the mean number
of infected neighbors is $b/(4 - 3b) m = 0.538411 m$,
consistent with Eq.\ (\ref{eq:probinfected}).

For $m = 4$, the results give the outgoing probabilities
in the construction of Kuulasmaa \cite{Kuulasmaa82}, in which a 
quenched double-directed
lattice showing all possible spreading probabilities is considered.
In this construction, one does not consider whether the neighbor is 
S or not, but draws outgoing directed bonds (with the proper probabilities) to 
all neighboring sites.  Then an epidemic is identified by following all
directed paths emanating from a given seed.

\begin{table}[htdp]
\caption{$P_k^{(m)}$ for the SIR model at $c = 0.176500$.
For comparison, the last row (4B) shows the binomial distribution
of bond percolation at $p_c = 1/2$, which is also at a critical point
but with much different $P_k^{(4)}$.}
\begin{center}
\begin{tabular}{|l|l|l|l|l|l|}
\hline
$m \setminus k$  &0 &1          &2              &3	   &4    \\
\hline

0	&1	  &		&               &	   &	 \\

1     &0.461589 &0.538411	&	        &          &	 \\

2     &0.300042 &0.323093	&0.376865       &          &	 \\

3     &0.222257 &0.233356	&0.251283       &0.293104  &     \\

4     &0.1765	  &0.1830288	&0.192169	&0.206931  &0.241371 \\
\hline
4B & 0.0625 & 0.25 & 0.375 & 0.25 & 0.0625 \\
\hline
\end{tabular}
\end{center}
\label{table:SIR}
\end{table}

\begin{figure}[htbp] 
%  figure placement: here, top, bottom, or page
\centering
\includegraphics[width=3in]{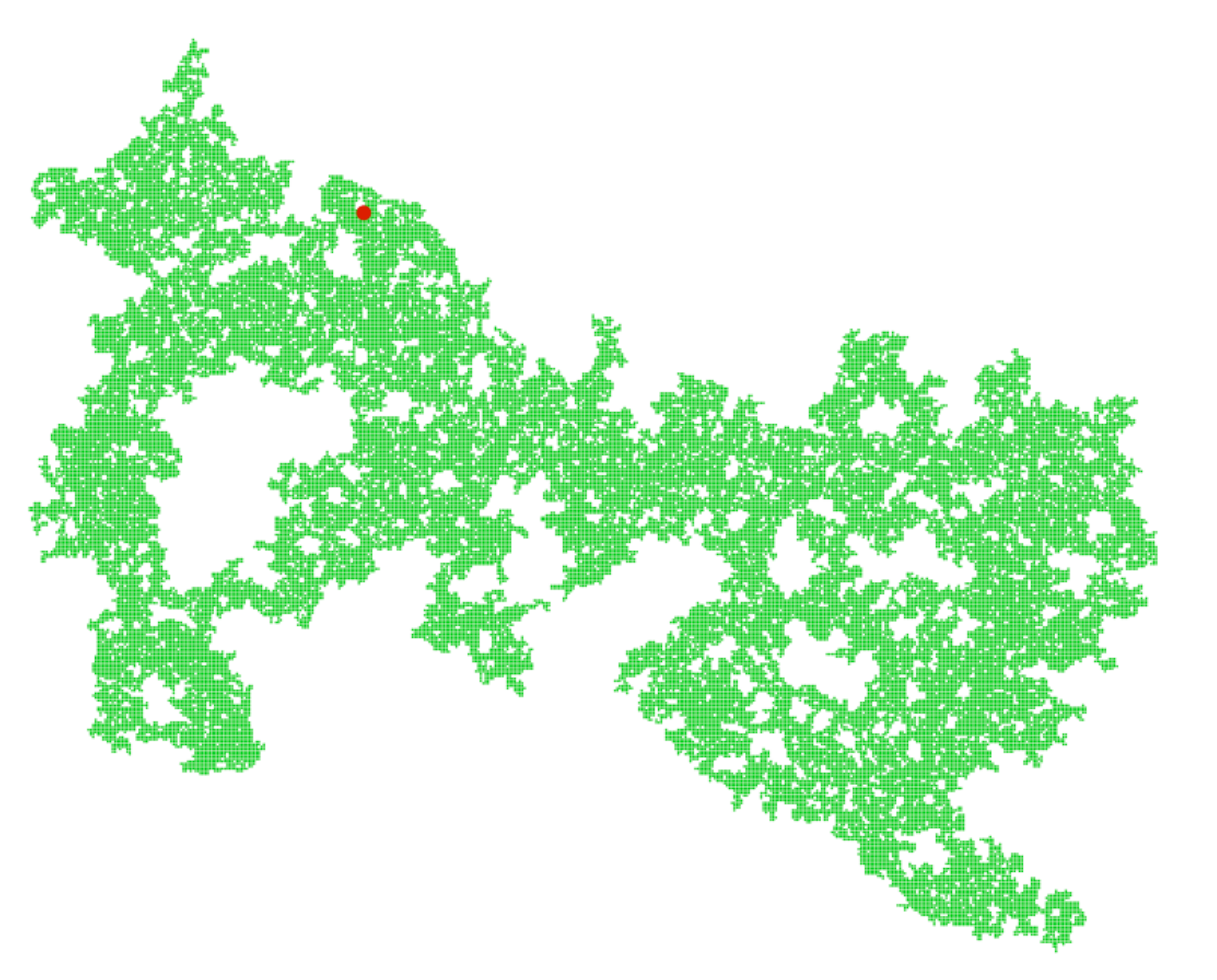} 
\caption{(Color online) Example of a completed bond percolation cluster of 51456 
wetted sites, generated at the threshold $p_c = 1/2$.  
The red (dark) dot marks the place where the cluster growth began.}
\label{fig:Perc51456}
\end{figure}

\begin{figure}[htbp] 
%  figure placement: here, top, bottom, or page
\centering
\includegraphics[width=3in]{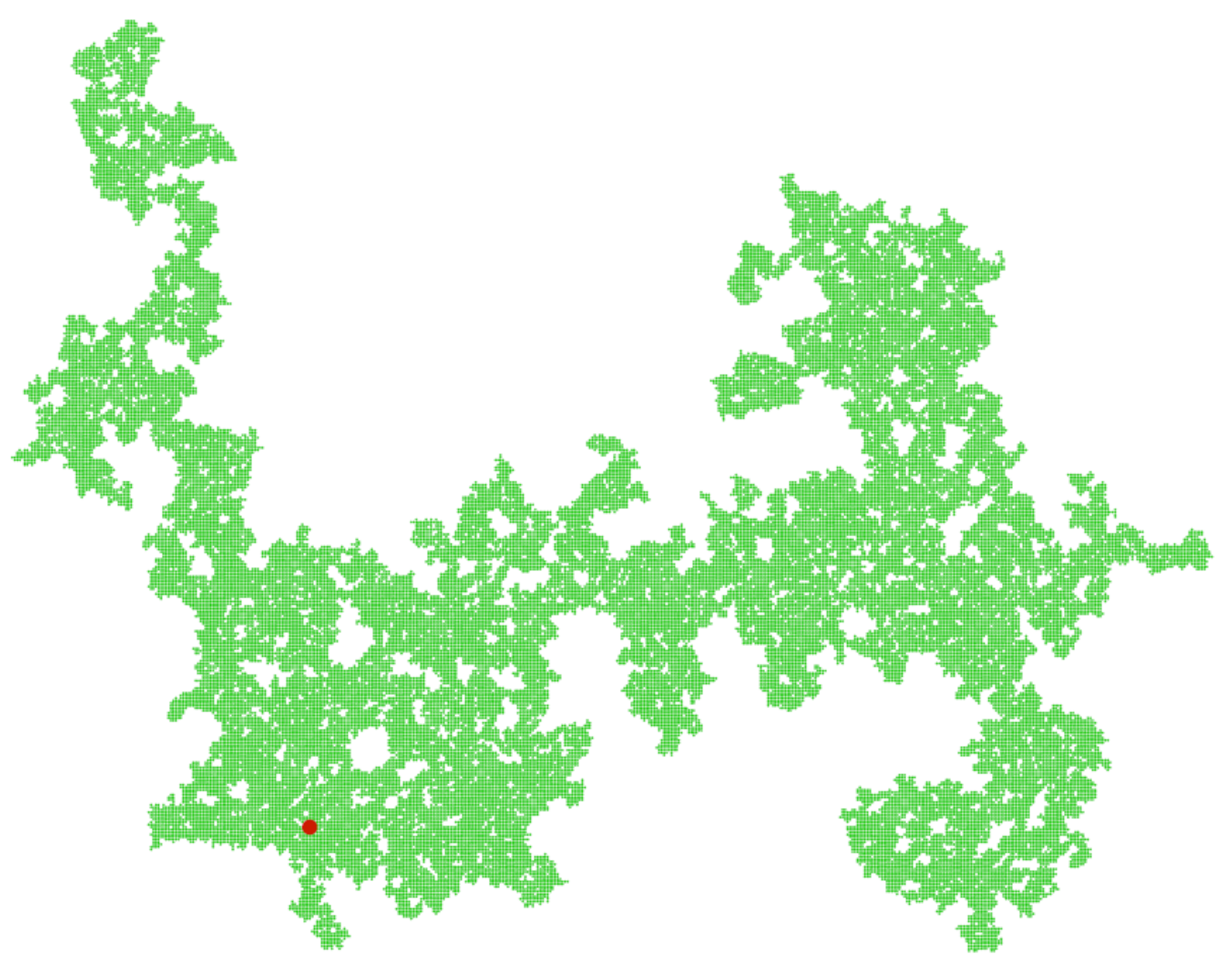} 
\caption{(Color online) Example of a completed SIR cluster of 51034 recovered sites, 
generated at the threshold $c_0 = 0.1765$.  
The red (dark) dot shows the location of origin of the infection.}
\label{fig:SIR51034}
\end{figure}

Thus, we see in general that in the SIR model at criticality there is an
enhancement in the number of neighbors that are infected, compared to the
case for dynamic percolation. Because of the correlations
in the infected neighbors in the SIR
model, it is necessary for the net neighbor infection probability to be higher
than the random percolation value of 1/2.

We note that for $c=1/5$, the value obtained by 
setting the probability of spreading (\ref{eq:probinfected})
equal to the bond percolation value $1/2$, $P_k^{(m)}$ is simply 
a uniformly distribution $P_k^{(m)} = 1/(m+1)$.  This behavior
is in sharp contrast to  (\ref{eq:binomial}) for $p = 1/2$ or 
indeed any value of $p$.

For the $z=4$ Bethe lattice at
its critical point $c=1/3$, the values of $P_k^{(4)}$ are
$5/15$, $4/15$, $3/15$, $2/15$ and $1/15$ for $k = 0$, 1, 
2, 3, and 4, respectively---trending in the opposite direction
as for the square lattice.

In Figs.\ \ref{fig:Perc51456} and  \ref{fig:SIR51034}  
we show actual pictures of critical clusters containing about 50,000
sites each from the percolation and SIR models, respectively.
On a local scale the SIR clusters appear slightly
denser but otherwise there is no apparent difference between the 
two critical clusters.

%--------------------------------------------
\section{Relation to other population biology models}
\label{sec:other}

The SIR model can be considered as a particular case of other
population biology models such as the  
Susceptible-Infected-Recovered-Susceptible (SIRS) model
\cite{deSouzaTome10,JooLebowitz04} and the Predator-Prey model 
\cite{ArashiroTome07}. 
The SIRS model describes an epidemic process without permanent
immunization and is defined by the following three processes:
S$\to$I, I$\to$R, and R$\to$S. The first two processes,
S$\to$I and I$\to$R, are the same as those for the SIR model,
as described in section 2,
and occur with rates $\beta$ and $\gamma$, respectively. 
The third process R$\to$S is spontaneous and occurs at rate $\alpha$.
The Predator-Prey model, in the epidemic language, is similar to the SIRS model except that the third
process, R$\to$S occurs with rate $\alpha n_\mathrm S/4$, where $n_\mathrm S$ is the number of
nearest-neighbor S sites.
In the epidemic language the following correspondence
is used: prey as S, predator as I, and an empty site as R.

The SIRS and the Predator-Prey models exhibit non-equilibrium 
phase transitions between 
an absorbing susceptible phase and active phase where the individuals
are continuously being infected. Their critical behavior belongs 
to the universality class of directed percolation for $\alpha\ne0$.
When $\alpha=0$ one recovers \cite{ArashiroTome07,deSouzaTome10}, 
from these two models the SIR model, which belongs to
the universality class of dynamic isotropic percolation,
as we have confirmed.

In the opposite regime, namely, when $\alpha$ is large enough
compared to $\beta$ and $\gamma$,
both the SIRS and the Predator-Prey models map 
\cite{deSouzaTome10,Kelly06} into the contact process (CP) 
\cite{Harris74} with a creation rate $\lambda=\beta/\gamma$.
The CP can be identified as 
the Susceptible-Infected-Susceptible (SIS) model
of epidemiological modeling \cite{Keeling08}.
The SIS model describes the dynamics of infection with no immunity and has two processes:  
S$\to$I with a rate $\beta n_\mathrm I/4$, and
I$\to$S  with rate $\gamma$.
Changing the time scale, these two processes can be described by the infection rate $\lambda =\beta / \gamma$ and 
recovery  rate equal to 1.
A site occupied by a particle in the CP model corresponds 
to an I site in the SIS model and an empty site to an S site.

The SIR and the SIS models correspond then to 
two extremal behaviors of the SIRS model with respect to immunity.
Let us consider the periods 
of time spent by an individual in the R state.
These periods of time are distributed around a mean value $T$,
which is proportional
to the inverse of the rate $\alpha$.
In the SIRS model $T$ is finite which means that
individuals have a partial immunization.
The SIR model can be understood as the SIRS model in
which $T\to\infty$,
meaning that an R individual has a lifelong immunity.
The SIS model, on the other hand, can be regarded as the SIRS 
model in which $T\to0$ implying that
an individual has no immunity or equivalently
that an infected individual becomes susceptible without
passing by the R state. 

Now we would like to give a comparison between the
critical parameters of the SIR and SIS models defined on
a square lattice. For this purpose,
it is convenient to use the parametrization
$b=\beta/(\alpha+\beta+\gamma)$, $c=\gamma/(\alpha+\beta+\gamma)$
and $a=\alpha/(\alpha+\beta+\gamma)$,
so that $b$, $c$ and  $a$
are interpreted as the probabilities of infection,
recovery and re-infection, respectively.
In this formulation there are just two independent parameters,
as $a+b+c=1$. 

For small values of $\beta$ and $\gamma$ compared to $\alpha$,
the critical line of the SIRS model can be obtained by 
means of the mapping into the SIS or CP \cite{deSouzaTome10,Kelly06}.
In two dimensions the critical value for the creation rate
for the CP is $\lambda_c=1.64874(4)$ 
\cite{Dickman99,SabagdeOliveira02,VojtaFarquharMast09}.  
Because $\lambda=\beta/\gamma$ and $b/c=\beta/\gamma$,
it follows that
the critical line is given by $b/c=\lambda_c=1.64874(4)$ as $a \to 1$.
If one extrapolates this line to $a=0$, 
$b+c=1$,  which corresponds to the SIR model
(that is, suppressing the parameter $a$),
one gets $(1-c)/c=\lambda_c$ or $c=1/(\lambda_c+1)\approx0.37754$
for the net probability that I$\to$S rather than
tries to infect one neighbor.
This value should be compared to the critical
value for the SIR process $c_0\approx0.1765$.
Alternatively, we may compare the infection probability
$b=1-c\approx 0.62247$ with the critical infection probability
for the SIR model $b_0=1-c_0\approx0.8235$.
It follows that epidemic spreading for
the SIR model occurs for a greater value of the infection
probability than the corresponding value for the SIS model.
This is expected because, in the SIS model, an S individual
can be infected multiple times.

%--------------------------------------------
\section{Conclusions}
\label{sec:conclusions}

We have provided further evidence that the asynchronous SIR model is in the
universality class of standard percolation through the behavior of the
cluster size distribution.  We showed that 
local correlations differ from those of standard percolation.  
By extensive numerical simulation of the
cluster size distribution, we have shown that the
transition in the SIR model defined on the square lattice occurs at $c_{0}=0.1765005(10)$,
consistent with, but much more precise than, previous work \cite{deSouzaTome10}.
This critical value of $c$ compares with $c = 1/3$ for the SIR model on
the 4-coordinated Bethe lattice, $0.37753$ for the contact process (or SIS model) on the
square lattice, and $0.407254$ for simultaneous infection of all neighboring S
sites on the square lattice (site percolation).
Having an accurate value of $c_0$ is useful for other studies of this critical state, such as
studying the scaling of the average cluster size with $L$ for finite systems
\cite{souzatome}.

{\it Note added in revision}.  While this paper was under review, a paper appeared
online 
\cite{NeriEtAl10} in which heterogeneity of the transmissibility in a square-lattice SIR model 
was also considered.  In that paper, the heterogeneity is formed by having classes of individuals
each with different fixed transmissibilities $\psi_i$ corresponding to having 
different fixed infections times.  In terms of the correlations we studied,
each of these classes of individuals produce a binomial distribution 
for $P_k^{(m)}$ given by Eq.\ (\ref{eq:binomial}) with $p$ replaced by
$\psi_i$, and by taking a set of classes it is possible for example to reproduce
the exponential infectivity distribution exactly.  Our critical transmissibility
$0.53841$ and its mean square deviation $\sigma^2 = 0.11575$
are consistent with the approximate criticality correlation these authors develop.
More details will be given in a future publication.
It should also be mentioned that that paper contains many references to
epidemiological systems where a lattice-based SIR model is relevant.

%---------------------------------------------------
\section{Acknowledgments}
This work was supported by the Brazilian agency CNPq and INCT of Complex Fluids (CNPq and FAPESP) (T.T.),
and supported in part by the U.\ S.\ National Science Foundation Grant No.\ DMS-0553487 (R.M.Z.).
%---------------------------------------------------
\section{Appendix} 

Consider that there are $m$ nearest-neighbor S sites to a given I site,
and four nearest neighbors total.
Then we define:

\begin{itemize}

\item $P^{(m)}_{n,k} =$ the probability that exactly $k$ distinct S 
sites are visited after $n$ trials on the I site, $k = 0, \ldots m$, 
for $n = 0, 1, 2, \ldots\infty$.

\end{itemize}
with $P^{(m)}_{0,0} = 1$; $P^{(m)}_{0,k} = 0$, $k = 1,\ldots,m$.
Given $p(n)$ defined in (\ref{eq:pn}), the net probability that $k$ 
sites of type S are visited is given by
\begin{equation}
P^{(m)}_{k} = \sum_{n=k}^\infty p(n) P^{(m)}_{n,k} 
=\sum_{n=k}^\infty c (1-c)^n P^{(m)}_{n,k}  \  ,
\label{eq:convolute}
\end{equation}
We can write recursion relations to find the $P^{(m)}_{n,k}$.  
For example, for $m = 1$ we have 
\begin{eqnarray}
P^{(1)}_{n,0} &=& \frac{3}{4} P^{(1)}_{n-1,0} \label{eq:P10} \\
P^{(1)}_{n,1} &=&  \frac{1}{4} P^{(1)}_{n-1,0} + P^{(1)}_{n-1,1} \ .
\label{eq:P11}
\end{eqnarray}
The first expression states that the probability that the S site was 
not visited at the $n$-th trial is
$3/4$ the probability it was not visited in the previous trial, 
as three out of four of the neighbors 
are not an S site.  Likewise, for the second expression, 
if the S was not visited in the previous trials,
then in one out of four times the S will be chosen for the first time in 
the $n$-th trial, while
if the S site was visited in some previous trial, then it will surely still 
have been visited in this trial.

To solve the recursion relations, we use the generating functions, 
defined for general $m$ by
\begin{equation}
G^{(m)}_k(x) = \sum_{n=k}^\infty P^{(m)}_{n,k} x^n  \  .
\end{equation}
A straightforward application to (\ref{eq:P10}-\ref{eq:P11}) yields
\begin{eqnarray}
G^{(1)}_0(x) &=& \frac{1}{1 - (3/4)x} \\
G^{(1)}_1(x) &=& \frac{x}{4 (1 - x)(1 - (3/4)x)} \ .
\end{eqnarray}
According to (\ref{eq:convolute}), we have simply
\begin{equation} 
P ^{(m)}_k = c G^{(m)}_k(1-c)  \  ,
\end{equation}
and we thus find
\begin{eqnarray}
P_0^{(1)} &=& \frac{4 c}{1 + 3 c} \\
P_1^{(1)} &=& \frac{ 1-c }{ 1 + 3 c}  \ ,
\end{eqnarray}
which satisfies $P_0^{(1)} + P_1^{(1)} = 1$.

Likewise, for $m=2$, we have
\begin{eqnarray}
P^{(2)} _{n,0} &=& \frac{2}{4} P^{(2)}_{n-1,0}\\
P^{(2)} _{n,1} &=& \frac{2}{4}  P^{(2)}_{n-1,0} +\frac{3}{4}  P^{(2)}_{n-1,1}\\
P^{(2)} _{n,2} &=& \frac{1}{4}  P^{(2)}_{n-1,1} + P^{(2)}_{n-1,2}
\end{eqnarray}
and similarly one can write the recursions for  $m = 3$ and 4.  In fact
one can summarize them for all $m$ by the general formulas
\begin{eqnarray}
P^{(m)} _{n,0} &=& \frac{4-m}{4} P^{(m)}_{n-1,0}\\
P^{(m)} _{n,k} &=& \frac{m+1-k}{4} P^{(m)}_{n-1,k-1}
+ \frac{4-m+k}{4} P^{(m)}_{n-1,k}
\end{eqnarray}
for $k=1,\ldots,m$, and derive a general recursion relation for the averaged quantities for all $m = 1, \ldots, 4$:
\begin{eqnarray}
P^{(m)} _{0} &=& \frac{4 c}{m+(4 - m)c} \label{eq:recursion0} \label{eq:recursion1}\\
P^{(m)} _{k} &=& \frac{(m-k+1)(1-c)}{m-k + (4 + k - m)c } P^{(m)} _{k-1}  \quad k = 1,...,m 
\label{eq:recursion2}
\end{eqnarray}

Using these formulas, we obtain the numerical results given in 
Table \ref{table:SIR}.  For $m = 4$ (the last row in that table), the
explicit formulas are:
\begin{eqnarray}
P^{(4)} _{0} &=& c \\
P^{(4)} _{1} &=& \frac{4 (1-c) c}{3 + c}\\
P^{(4)} _{2}  &=& \frac{12 (1-c)^2 c}{(3 + c)(2+2 c)}\\
P^{(4)} _{3} &=& \frac{24 (1-c)^3 c}{(3 + c)(2+2 c)(1 + 3 c)}\\
P^{(4)} _{4} &=& \frac{6 (1-c)^4}{(3 + c)(2+2 c)(1 + 3 c)}
\end{eqnarray}

Note that it is possible to write explicit formulas for all the 
$P^{(m)}_{n,k}$, such as
\begin{equation}
P^{(3)}_{n,3} = \frac{4^n - 3(3^n) + 3(2^n) - 1}{4^n}  \ ,
\end{equation}
where the coefficients in some cases are related to Stirling numbers of the second kind.
Also, we can generalize  (\ref{eq:recursion1}-\ref{eq:recursion2})  to all coordination numbers $z$ simply
by replacing all 4's in those equations by $z$'s.  Then one can show that 
\begin{equation}
P_m^{(m)} + P_{m-1}^{(m)} + \ldots + P_{m-n+1}^{(m)} = \frac{n(1-c)}{z c}P_{m-n}^{(m)} \ ,
\end{equation}
for $n = 1, \ldots, m$, 
and taking $n=m$, one can verify directly that $\sum_{k=0}^m P_k^{(m)} = 1$.
%---------------------------------------------------

\end{document}